\documentclass[12pt]{article}
\usepackage[utf8]{inputenc}
\usepackage[top=50pt,bottom=50pt,left=68pt,right=66pt]{geometry}
\usepackage{amsmath}
\usepackage{bm}
\usepackage{booktabs}
\usepackage{amssymb}
\usepackage[title]{appendix}
\usepackage{mathrsfs}
\usepackage{cite}
\usepackage{float}
\usepackage{xcolor}
\usepackage{multirow}
\usepackage{graphicx,caption,subcaption}
\usepackage[space]{grffile}
\usepackage{ upgreek }

\renewcommand{\arraystretch}{1.3}
\definecolor{darkblue}{rgb}{0.1,0.1,.7}
\usepackage[breaklinks=true,linktocpage=true,
colorlinks=true,urlcolor=darkblue,linkcolor=darkblue,
citecolor=darkblue,pdfpagelabels=true,hypertexnames=true,
plainpages=false,naturalnames=false,]{hyperref}

\usepackage{datetime}
\newdateformat{monthyeardate}{%
  \monthname[\THEMONTH], \THEYEAR}
\date{\monthyeardate\today}

\def\cl{{\cal L}}

\def\cp{{\cal P}}

\def\cu{{\cal U}}

\begin{document}

\renewcommand{\arraystretch}{1.3}
\thispagestyle{empty}

{\hbox to\hsize{\vbox{\noindent\monthyeardate\today}}}

\noindent
\vskip2.0cm
\begin{center}

{\Large\bf Strong primordial inhomogeneities of axion-like field in Einstein--Gauss--Bonnet gravity}

\vglue.3in

M.A. Krasnov${}^{a,b}$,\,
D.Z. Berkimbayev${}^{c}$,\,
A. Addazi${}^{d,e}$,\,
Y. Aldabergenov${}^{f,c}$,\,
M.Y. Khlopov${}^{g}$
\vglue.1in

${}^a$~{\it National Research Nuclear University MEPhI, Moscow, 115409, Russia}\\
${}^b$~{\it Research Institute of Physics, Southern Federal University, Rostov-on-Don, 344090, Russia}\\
${}^c$~{\it Al-Farabi Kazakh National University, Almaty, 050040, Kazakhstan}\\
${}^d$~{\it Institute of Astronomy and Astrophysics, School of Mathematics and Physics, Anqing Normal University, Anqing 246133, China}\\
${}^e$~{\it Laboratori Nazionali di Frascati INFN, Rome, 344090, Italy}\\
${}^f$~{\it Department of Physics, Fudan University, Shanghai, 200433, China}\\
${}^g$~{\it Virtual Institute of Astroparticle physics, Paris, 75018, France}\\

\vglue.1in

\end{center}

\vglue.3in

\begin{center}
{\Large\bf Abstract}
\vglue.2in
\end{center}

In this paper we consider complex scalar field with Mexican hat potential coupled with Gauss-Bonnet scalar resulting in restored $U(1)$ symmetry at the stage of inflation, corresponding to the modern cosmological horizon. In our model, the phase transition takes place during inflation, avoiding large-scale axion isocurvature perturbations. Our model provides strong primordial inhomogeneities on small scales and ultra-light axion-like particle as either a fuzzy dark matter candidate or a candidate for inhomogeneous dark energy. We also study the possibility of black hole formation and gravitational wave background in this scenario. We find that in the case when the symmetry breaking field does not backreact on inflation, black hole production is not feasible. However, under specific conditions, gravitational waves fitting the NANOGrav band can be obtained.


\section{Introduction}

Among modern variety of modified gravity theories there is a Gauss-Bonnet (GB) modification of general relativity, which was initially proposed in papers by Lanczos \cite{d8c23c97-1e7d-30cf-af7c-c0828c0d0c95, Lanczos1932}. This modification of gravity does not introduce more than two derivatives of the metric in the equations of motion, which makes it free of Ostrogradsky instabilities \cite{Ostrogradsky:1850fid}. On its own, the GB term is topological, but when coupled to scalar fields, it modifies the field equations \cite{Satoh_2008, Guo_2007, Cartier_2001} (for an extensive review on GB gravity, see e.g. \cite{Fernandes:2022zrq}). When a scalar field that couples to the GB term transforms under some symmetry, the GB term can create new effective minima which spontaneously break or restore the symmetry \cite{Aldabergenov_2025,Krasnov:2025ssc}. These effective minima depend on the Hubble parameter, which creates interesting scenarios during inflation: as the inflationary Hubble parameter slowly varies with time, this leads to dynamical nature of the effective minima, and the possibility of symmetries being broken or restored in the middle of inflation. Based on this idea, we consider axions, or axion-like particles (ALPs), created during inflation from such a mechanism.

An ALP is a pseudo-Nambu-Goldstone boson which, despite its small mass, has become a popular dark matter candidate \cite{Sikivie_2008, Di_Luzio_2020, dobrich2025experimentstesthypothesissolar}. The reason behind this is the misalignment mechanism~--- in the vicinity of potential's minimum, it behaves like pressureless matter. Furthermore, because of its small mass, its wavelength is large enough to make it possible for it to form structures. For an extensive review on axion cosmology see \cite{OHare:2024nmr}.

In this paper we propose a novel ALP cosmological scenario that provides enhanced ALP-induced inhomogeneities at small scales while avoiding large-scale isocurvature problem. In our particular scenario, the model provides a fuzzy dark matter \cite{PhysRevLett.85.1158} candidate (for a review on fuzzy dark matter see \cite{eberhardt2025ultralightfuzzydarkmatter, Schive_2026}). Currently, fuzzy dark matter is constrained in a broad mass range, see e.g. \cite{sipple2025fuzzydarkmatterconstraints, 2024PhRvD.110l3532L}.



The idea of our study follows from Refs. \cite{Rubin:2000dq,Rubin:2001yw,Khlopov:2002yi,Khlopov:2004sc}, where a mechanism of the formation of massive domain walls after inflation is considered, based on a global $U(1)$ symmetry breaking model,
\begin{equation}\label{L_SSB}
    \sqrt{-g}^{-1}\cl=\tfrac{1}{2}R-\partial \Phi\partial\Phi^*-\tfrac{\lambda}{4}(f^2-2\Phi\Phi^*)^2~,
\end{equation}
where $\Phi$ is a complex scalar of Peccei--Quinn type \cite{PhysRevLett.38.1440}, $\lambda$ is the quartic coupling, and $f$ is the ALP decay constant acting as a vacuum expectation value (VEV) parameter, $\langle\Phi\rangle=f/\sqrt{2}$. Unless otherwise mentioned, \textcolor{black}{reduced} Planck units are used, $M_{\rm Pl}=1$. Around the symmetry breaking vacuum, $\Phi$ can be parametrized as
\begin{equation}
    \Phi=\tfrac{1}{\sqrt{2}}\phi e^{-i\theta}~,
\end{equation}
where $\theta$ is the ALP. After inflation, instanton effects generate a potential for the ALP,
\begin{equation}
    \delta V(\theta)=\Lambda^4(1-\cos\theta)~,
\end{equation}
where the parameter $\Lambda$ is much smaller than the inflationary scale.

In \cite{Rubin:2000dq,Rubin:2001yw,Khlopov:2002yi,Khlopov:2004sc} it was assumed that the symmetry is already broken during (observable) inflation, so that the massless ALP is present during inflation. As inflation offers a quasi-de Sitter background for the ALP, the latter fluctuates with the amplitude \cite{Starobinsky:1986fx,Nambu:1987ef,Nakao:1988yi}
\begin{equation}\label{axion_fluct}
    \delta\theta\simeq \frac{H}{2\pi f}~,
\end{equation}
over each Hubble time $1/H$. This leads to the appearance of domains with $\theta>\pi$ surrounded by domains with $\theta<\pi$. After inflation ends and $H$ drops below a certain value, the ALP $\theta$ relaxes to one of its minima: in the domains with $\theta>\pi$, $\theta$ relaxes to $\theta=0$, and in the domains with $3\pi>\theta>\pi$, to $\theta=2\pi$, and so on.

The idea is to consider a time-dependent VEV $f(t)$, so that at the beginning of inflation $f=0$ and the symmetry is restored (${\rm Re}\Phi$ and ${\rm Im}\Phi$ are two massive scalars in this case), and at some point during inflation $f$ becomes non-zero, at which point the ALP is generated around non-zero VEV of $\phi$, and starts to fluctuate according to \eqref{axion_fluct}. This can be achieved by, e.g., the GB coupling of $\Phi$. Idea of such a cosmological scenario is proposed in \cite{Redi_2023}, although without a specific inflationary mechanism.

In our analysis, we also study the possibility of formation of primordial black holes (PBHs) and gravitational waves via collapsing ALP domain walls in a model with delayed $U(1)$ symmetry breaking during inflation.

The paper is organized as follows: in section II we discuss how GB term could provide symmetry breaking during inflationary epoch, section III is dedicated to the clarification of the problem we consider, in section IV we study quantum fluctuations of the field and derive initial conditions for classical motion of the radial part of the complex scalar field, in section V we calculate power spectrum of ALP fluctuations and their characteristic scale, in section VI we discuss potential cosmological scenarious within our model. We also discuss PBH formation and gravitational wave background in subsections of section VI.

\section{A model based on Gauss--Bonnet-induced symmetry restoration}

In our model the Lagrangian \eqref{L_SSB} is supplemented by the GB term
\begin{equation}\label{L_GB}
\sqrt{-g}^{\,-1}\mathcal{L}_{\rm GB} = -\tfrac{1}{8}\xi(|\Phi|) R_{\rm GB}^{2}~,
\end{equation}
where
\begin{equation}
    R^2_{\rm GB}\equiv R^2-4R_{\mu\nu}R^{\mu\nu}+R_{\mu\nu\rho\sigma}R^{\mu\nu\rho\sigma}~,
\end{equation}
and $\xi(|\Phi|)$ is a function of the absolute value $|\Phi|=\phi/\sqrt{2}$. This leads to the effective potential
\begin{equation}
    V_{\rm eff}(\phi)=V(\phi)+\tfrac{1}{8}\xi(\phi)R^2_{\rm GB}~,
\end{equation}
where $V$ is the potential from \eqref{L_SSB}, $V=\tfrac{1}{4}\lambda(f^2-\phi^2)^2$, with the negative mass-squared around the symmetric point. The GB contribution can shift the effective mass-squared to positive values at the beginning of inflation,
\begin{equation}\label{eff_mass}
    V_{,\phi\phi}+\tfrac{1}{8}\xi_{,\phi\phi}R_{\rm GB}^2>0~,
\end{equation}
so that the ALP isocurvature problem is avoided. If the second term in \eqref{eff_mass} decreases with time during inflation, at some point the effective mass-squared becomes negative, and the symmetry breaking occurs. After this point the ALP starts fluctuating, and the scenario develops as in the previous section, except for the presence of gradually decreasing GB term.

The simplest choice is
\begin{equation}
    \xi=\tfrac{1}{6}\alpha\phi^2~,
\end{equation}
with some parameter $\alpha$.

\section{Problem formulation}
 The starting point is to study the evolution of VEV $f_{\rm eff}(t)$. 
Let us expand the potential:
\begin{multline}
    V_{\rm eff}(\phi)=V(\phi)+\tfrac{1}{8}\xi(\phi)R^2_{\rm GB}=\tfrac{1}{4}\lambda(f^2-\phi^2)^2+3\xi H^2(H^2 + \Dot{H})=\\
    =\frac{1}{4}\lambda \phi^4+\frac{1}{2}\left( \alpha H^2(H^2+\Dot{H})-\lambda f^2\right) \phi^2+\frac{\lambda}{4}f^4~.
\end{multline}
We can introduce $f_{\rm eff}(t)$ as the solution of the following equation,
\begin{equation}
    \cfrac{d V_{\rm eff}}{d\phi}\bigg|_{\phi=f_{\rm eff}(t)}=0~,
\end{equation}
which could be written as
\begin{equation}\label{Extremum}
   f_{\rm eff}(t)[ \lambda f_{\rm eff}(t)^2+\alpha H^2(H^2+\Dot{H})-\lambda f^2]=0~.
\end{equation}
The point $f_{\rm eff}(t)=0$ corresponds to the symmetric point, while for its non-zero value, Eq.\eqref{Extremum} is solved as
\begin{equation}\label{f(t)_equation}
    f_{\rm eff}(t)=f\sqrt{1-\cfrac{\alpha H^2}{\lambda f^2}(H^2+\Dot{H})}~.
\end{equation}

We assume that the evolution of $H$ is governed by some inflationary model, and which is not significantly affected by the dynamics of $\phi$. Let us write the whole action, with the inclusion of the inflaton $\chi$, as follows,
\begin{equation}
    S=\int d^4x\sqrt{-g}\left[ \tfrac{1}{2}R-\partial \Phi\partial\Phi^*-\tfrac{\lambda}{4}(f^2-2\Phi\Phi^*)^2 -\tfrac{1}{8}\xi(|\Phi|) R_{GB}^{2} + \mathcal{L}_{inf}(\chi)\right]~.
\end{equation}
We adopt Starobinsky-like model for $\chi$ \cite{STAROBINSKY198099}:
\begin{equation}
    \mathcal{U}(\chi)=\cfrac{1}{2}M^2\left(1-e^{-\chi}\right)^2~.
\end{equation}
The inflaton has the following equation of motion (dot represents derivative with respect to cosmic time $t$):
\begin{equation}\label{InflatonChi}
    \Ddot{\chi}+3H\Dot{\chi}+\mathcal{U}_{,\chi}=0~,
\end{equation}
and Friedmann equations are
\begin{align}
    3H^2(1-\Dot{\xi}H)=\cfrac{1}{2}(\Dot{\phi}^2 + \Dot{\chi}^2)+\mathcal{U}(\chi)+V_{\rm eff}(\phi)~,\\
    2 \Dot{H}(1 - \Dot{\xi}H)-\Ddot{\xi}H^2 + \Dot{\xi}H^3 + \Dot{\phi}^2 + \Dot{\chi}^2 = 0~.
\end{align}

Provided that initially Eq. \eqref{Extremum} has only the $f_{\rm eff}(t)=0$ solution (Eq. \eqref{f(t)_equation} is imaginary), the moment of time at which non-zero VEV develops is found from the zero crossing,
\begin{equation}\label{VeV}
    f_{\rm eff}(t)=f\sqrt{1-\cfrac{\alpha H^2}{\lambda f^2}(H^2+\Dot{H})} =0~,
\end{equation}
after which $f_{\rm eff}(t)$ becomes real and positive. Combining equations above (\ref{InflatonChi}-\ref{VeV}) we can find the symmetry breaking time (and the corresponding e-fold number), which can be controlled by the choice of the parameters.

\begin{figure}[H]
	\begin{center}
\includegraphics[width=0.8\textwidth]{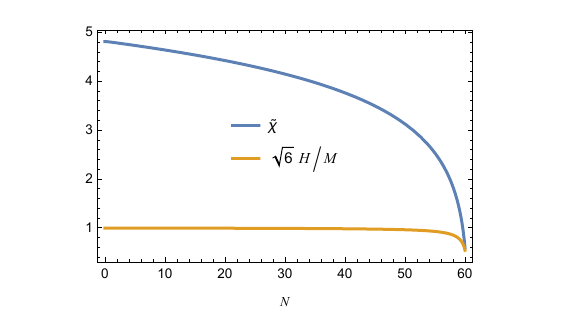}
	\end{center}
\caption{Numerical solution to the equation of motion of the inflaton field $\chi$ from system \eqref{EOM_inf_phi_N}, which is then utilized to obtain numerical dependence of Hubble parameter on e-fold.}
	\label{Inflaton}
\end{figure}

It is convenient to use the forward e-fold number (satisfying $H=\dot N$) instead of the cosmic time $t$. Let us also introduce the Hubble slow-roll parameter $\epsilon\equiv-H_{,N}/H$, and the GB slow-roll parameter $\omega\equiv \xi_{,N}H^2$. This leads to the equations of motion
\begin{align}
    \chi_{,NN}+(3-\epsilon)\chi_{,N}+\cu_{,\chi}H^{-2} &=0~,\label{EOM_inf_chi_N}\\
    \phi_{,NN}+(3-\epsilon)\phi_{,N}+V_{,\phi}H^{-2}+3\xi_{,\phi} H^2(1-\epsilon) &=0~,\label{EOM_inf_phi_N}
\end{align}
from which we obtain numerical solution for the inflaton $\chi$ (for effectively single-field inflation). We demonstrate the solution in Fig.~\eqref{Inflaton}. The evolution of the massless axion $\theta$ will be discussed in the next section.~\footnote{Pre-, post-SSB fields}

\begin{figure}[H]
	\begin{center}
\includegraphics[width=0.8\textwidth]{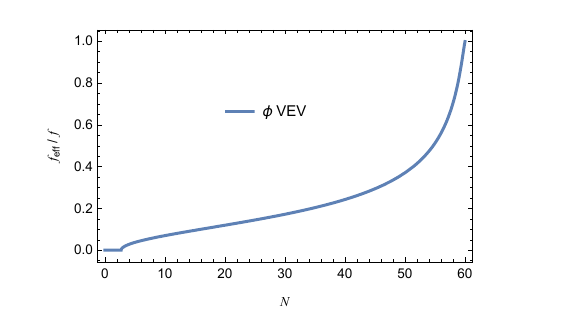}
	\end{center}
\caption{Numerical evolution of the effective VEV \eqref{VeV_N}. In the region, where expression within the square root is negative, $f_{\rm eff}$ is set to zero, as it is the only critical point. We define the e-fold at which $f_{\rm eff}$ becomes positive as $N_c$.}
	\label{VEV}
\end{figure}

The first Friedmann equation is
\begin{equation}
    3(1-\xi_{,N}H^2)H^2-\tfrac{1}{2}H^2(\chi_{,N}^2+\phi_{,N}^2)-V(\phi)-\cu(\chi)=0~,\label{H_constraint_full}
\end{equation}
which is quartic in $H$. From \eqref{H_constraint_full} we find
\begin{equation}
    H^2=\frac{2(V+\cu)}{3-\tfrac{1}{2}(\chi_{,N}^2+\phi_{,N}^2)}\bigg\{1+\sqrt{1-\frac{4\xi_{,N}(V+\cu)}{3[1-\tfrac{1}{6}(\chi_{,N}^2+\phi_{,N}^2)]^2}}\bigg\}^{-1}~.
\end{equation}
The effective VEV can be written as
\begin{equation}\label{VeV_N}
    f_{\rm eff}(N)=f\sqrt{1-\frac{\alpha H^4}{\lambda f^2}(1-\epsilon)}~.
\end{equation}
We show the numerical evolution of the effective VEV \eqref{VeV_N} in Fig.~\eqref{VEV}. Here we would like to emphasize a subtlety related to quantum diffusion effects around $\phi=0$. Namely, $f_{\text{eff}}>0$ does not immediately imply the breaking of the $U(1)$ symmetry in the classical sense. This is because around $\phi=0$ there can be a period of time when quantum fluctuations of $\phi$ are dominant over the classical motion, and the expectation value $\langle\phi\rangle$ is zero. Therefore we will consider the symmetry to be broken only when the quantum effects subside. We will present a quantitative analysis regarding this point later in the text.

To avoid backreaction of $\phi$ on inflation, we impose
\begin{equation}
    \lambda f^4 \ll M^2~,
\end{equation}
and \cite{Aldabergenov_2025}
\begin{equation}\label{no_backreaction_condition}
     |\omega|,|\omega_{,N}|\ll \epsilon\ll 1~,
\end{equation}
i.e., the GB slow-roll parameter and its rate of change are subdominant to $\epsilon$. It allows for effectively single-field inflation (Starobinsky-type inflation in this case, but any other slow-roll single-field model can be considered). In particular, the Hubble function becomes $H\simeq \sqrt{\cu(\chi)/3}$.

Under our assumptions, the following parameter values can be chosen,

\begin{equation} \label{parameters}
    \lambda=10^{-10.2}~,~~f=AM~,~~ \alpha=\cfrac{b}{M^2}~,~~A=125\cdot10^3~,~~b=37.25~.
\end{equation}
The values in \eqref{parameters} are chosen to fulfill the subdominancy of the PQ field. With this choice, the e-fold time of the symmetry breaking is $N\approx 30$ (after the classicalization, see below). Note that this choice of the parameters is not unique, and taken as an example.

\section{Stochastic evolution of the ALP}

Here we apply stochastic formalism to describe fluctuations of the ALP $\theta$ during slow-roll inflation. In stochastic formalism we split $\theta$ as
\begin{equation}
    \theta(t,x)=\theta_c(t)+\theta_q(t,x)~,
\end{equation}
where $\theta_c$ is the classical part -- a collection of long wavelength modes, and $\theta_q$ is the quantum part -- a collection of short wavelength modes. The equations of motion are often linearized in $\theta_q$, but non-linearities in $\theta_c$ are retained. The difference with the standard background-perturbation split $\theta=\bar\theta(t)+\delta\theta(t,x)$, is in the Fourier transform of $\delta\theta$ and $\theta_q$,
\begin{align}
    \delta\theta &=\int\frac{d^3k}{(2\pi)^{3/2}}(e^{-i{\bf kx}}\delta\theta_k \hat a_k+{\rm h.c.})~,\\
    \theta_q &=\int\frac{d^3k}{(2\pi)^{3/2}}W\Big(\frac{k}{k_\sigma}\Big)(e^{-i{\bf kx}}\delta\theta_k \hat a_k+{\rm h.c.})~,\label{Fourier_theta_q}
\end{align}
where $W$ is a window function, and $k_\sigma\equiv\sigma aH$ (inverse coarse-graining scale) with $\sigma\ll 1$. The window function keeps short wavelength modes modes with $k\geq k_\sigma$. The simplest choice of $W$ is the step function $W(X)=\Theta(X-1)$, so that $\dot W(X)=\delta(X-1)\dot X$. If the windows function is constant (and normalized as $W=1$), $\theta_c=\bar\theta$ and $\delta\theta=\theta_q$. Time depdendence of $W$ in \eqref{Fourier_theta_q} produces stochastic noise in the equation of motion for the classical field $\theta_c$. On the other hand, since $\theta_q$ includes $\delta\theta_k$ in its Fourier transform, $\delta\theta_k$ must be known in order to derive statistics of the stochastic noise. But to solve the EOM for the linearized perturbation $\delta\theta_k$, background ($c$-fields) evolution must be derived first, which is affected by the noise. This problem can be resolve if we know the initial conditions (e.g., $\theta_c$ and $\dot\theta_c$ and other relevant fields) at the moment when the system enters noise-dominated regime (quantum diffusion). Numerically, one can discretize time, and at the first step use the aforementioned initial conditions, neglect the noise in the $c$-fields equation, and use this background solution to find the evolution of $\delta\theta_k$. This gives us the noise statistics, which can be used at the next time step, to solve the stochastic equations for the background fields, now in the presence of the noise. The process can be repeated at each time step, and with some (preferably large) number of realizations, in order to build a probability distribution for the field values of $\theta_c$ at a given time.

Slow-roll inflation can simplify the equations and enable us to find some analytical results, for example for the Fourier mode ($\delta\theta_k$) solution and the associated noise statistics. This will be shown below.

Let us start by considering the full Lagrangian, including the inflaton sector,
\begin{equation}\label{L_full_stoch}
    \sqrt{-g}^{-1}\cl=\tfrac{1}{2}R-\tfrac{1}{8}\xi(\phi)R^2_{\rm GB}-\tfrac{1}{2}(\partial \chi\partial\chi+\partial \phi\partial\phi+\phi^2\partial \theta\partial\theta)-\cu(\chi)-\tfrac{\lambda}{4}(f^2-\phi^2)^2~.
\end{equation}
In order to treat the perturbations or $q$-fields rigorously, we should also introduce metric perturbations. For convenience, we choose spatially flat gauge, where metric elements and inverse metric elements can be written as (up to second order in perturbations)
\begin{gather}
\begin{gathered}\label{alpha_bete_metric}
    g_{00}=-(1+\alpha)^2+a^{-2}\partial_i\beta\partial^i\beta~,~~~g_{0i}=\partial_i\beta~,~~~g_{ij}=a^2\delta_{ij}~,\\
    g^{00}=-(1-2\alpha+3\alpha^2)~,~~~g^{0i}=a^{-2}(1-2\alpha)\partial^i\beta~,~~~g^{ij}=a^{-2}\delta^{ij}-a^{-4}\partial^i\beta\partial^j\beta~,
\end{gathered}
\end{gather}
where the spatial indices $i,j$ can be raised and lowered by the Cronecker delta. The variable $\alpha$ is perturbation of the lapse function, and $\partial_i\beta$ is the shift vector. Spatial perturbation is absent in this gauge (tensor perturbations are ignored in this analysis).

Background equation for $\bar\theta$ is particularly simple,
\begin{equation}\label{theta_background_EOM}
    \ddot{\bar\theta}+\Big(3H+2\frac{\dot\phi}{\phi}\Big)\dot{\bar\theta}=0~,
\end{equation}
as there is no potential for it during inflation. Provided that
\begin{equation}\label{SR_phi}
    \Big|\frac{2\dot\phi}{3H\phi}\Big|=\Big|\frac{2\phi_{,N}}{3\phi}\Big|\ll 1~,
\end{equation}

Eq. \eqref{theta_background_EOM} can be solved, in terms of the e-fold time $N$, as
\begin{equation}\label{theta_USR_sol}  \bar\theta(N)\simeq\bar\theta(N_{\rm in})-\tfrac{1}{3}\bar\theta_{,N}(N_{\rm in})e^{3(N_{\rm in}-N)}~,
\end{equation}
where $N_{\rm in}$ is some initial time (after the symmetry breaking), and $\bar\theta(N_{\rm in})$ and $\bar\theta_{,N}(N_{\rm in})$ are initial field value and velocity.
We would check the condition \eqref{SR_phi} later in the text. 

\subsection{Linear perturbation theory for the ALP}

Next, we derive linearized EOM for $\delta\theta_k$. We perturb only the ALP, $\theta=\bar\theta+\delta\theta$, while keeping $\chi$ and $\phi$ as background fields, assuming that inflation is driven by $\chi$, and the PQ sector does not backreact. As mentioned before, this is guaranteed if
\begin{equation}\label{SR_epsilon_omega}
    |\omega|,|\omega_{,N}|\ll \epsilon\ll 1~,
\end{equation}
where $\omega\equiv\dot\xi H=\xi_{,N}H^2$.

By using the metric \eqref{alpha_bete_metric} and perturbting the ALP, we get the second-order Lagrangian
\begin{align}
\begin{aligned}\label{L_2_pert}
    a^{-3}\cl^{(2)}= &-\big[3H^2(1-2\omega)-\tfrac{1}{2}(\dot\chi^2+\dot\phi^2+\phi^2\dot{\bar\theta}^2)\big]\alpha^2-(2-3\omega)H\alpha\frac{\partial_i^2}{a^2}\beta\\
    &-\tfrac{1}{2}\phi^2\Big(2\dot{\bar\theta}\alpha\delta\dot\theta-\delta\dot\theta^2-2\dot{\bar\theta}\delta\theta\frac{\partial_i^2}{a^2}\beta-\delta\theta\frac{\partial_i^2}{a^2}\delta\theta\Big)~.
\end{aligned}
\end{align}
The metric perturbations can be eliminated as
\begin{equation}\label{alpha_eq}
    \alpha=\frac{\phi^2\dot{\bar\theta}\delta\theta}{(2-3\omega)H}~,
\end{equation}
and
\begin{equation}\label{beta_eq}
    -\frac{\partial_i^2}{a^2}\beta=\frac{\phi^2\dot{\bar\theta}\delta\dot\theta}{(2-3\omega)H}+[6H^2(1-2\omega)-\dot\chi^2-\dot\phi^2-\phi^2\dot{\bar\theta}^2]\frac{\phi^2\dot{\bar\theta}\delta\theta}{(2-3\omega)^2H^2}~.
\end{equation}
Pluggin them back in \eqref{L_2_pert} yields
\begin{align}
\begin{aligned}\label{L_2_delta_theta}
    a^{-3}\cl^{(2)}= &~\frac{\phi^2}{2}\Big(\delta\dot\theta^2+2\delta\theta\frac{\partial_i^2}{a^2}\delta\theta\Big)+a^{-3}\partial_t\Big[\frac{a^3\phi^4\dot{\bar\theta}^2}{2(2-3\omega)H}\Big]\delta\theta^2\\
    &-\big[3H^2(1-2\omega)-\tfrac{1}{2}(\dot\chi^2+\dot\phi^2+\phi^2\dot{\bar\theta}^2)\big]\frac{\phi^4\dot{\bar\theta}^2\delta\theta^2}{(2-3\omega)^2H^2}~.
\end{aligned}
\end{align}
By using \eqref{theta_background_EOM}, \eqref{SR_phi}, \eqref{SR_epsilon_omega}, as well as $|\phi_{,NN}/\phi|\ll 1$, from the Lagrangian \eqref{L_2_delta_theta} we obtain approximated EOM for the Fourier modes $\delta\theta_k$,
\begin{equation}\label{d_theta_k_eq}
    \delta\ddot\theta_k+3H\delta\dot\theta_k+\Big(\frac{k^2}{a^2}+3\phi^2\dot{\bar\theta}^2\Big)\delta\theta_k\simeq 0~.
\end{equation}
Or, in terms of $u_k\equiv a\phi\delta\theta_k$ and the e-fold time,
\begin{equation}\label{u_k_eq_N}
    u_{k,NN}+u_{k,N}+\Big[\frac{k^2}{a^2H^2}-2+3\phi^2\bar\theta_{,N}^2(N_{\rm in})e^{6(N_{\rm in}-N)}\Big]u_k\simeq 0~,
\end{equation}
where \eqref{theta_USR_sol} was used, and $N_{\rm in}$ is some initial time, for example soon after the critical time $N_c$.
Suppose that~\footnote{Under slowly rolling $\phi$, as in \eqref{SR_phi}, the condition \eqref{theta_N_condition} means slowly changing canonical ALP $\phi\bar\theta$. This is justified because during the first e-folds after the symmetry breaking, the only contribution to the velocities of $\phi$ and $\bar\theta$ is quantum diffusion, so that $(\phi\bar\theta)_{,N}\sim H/(2\pi)\ll 1$.}
\begin{equation}\label{theta_N_condition}
    \phi^2(N_{\rm in})\bar\theta_{,N}^2(N_{\rm in})\ll 1~.
\end{equation}
Then, as time passes $(N>N_{\rm in})$, the factor $e^{6(N_{\rm in}-N)}$ will lead to further suppression of the last term of \eqref{u_k_eq_N}, as long as $\phi^2(N)$ does not grow exponentially. Therefore, if \eqref{theta_N_condition} holds, one can ignore this term, so that \eqref{u_k_eq_N} can be solved as
\begin{equation}\label{dtheta_SR_solution}
    \delta\theta_k=\frac{u_k}{a\phi}\simeq\frac{e^{-ik\tau}}{a\phi\sqrt{2k}}\Big(1+i\frac{aH}{k}\Big)~,
\end{equation}
where $\tau\simeq -1/(aH)$ is the conformal time.

\subsection{Stochastic formalism for the ALP}

In order to obtain stochastic Klein--Gordon or Langevin equations for the ALP, we use $\theta=\theta_c+\theta_q$ and leave $\chi$ and $\phi$ unperturbed as before (treated as $c$-fields). The metric perturbations $\alpha$ and $\beta$ should now be treated as $q$-fields. The Lagrangian \eqref{L_full_stoch} is then expanded up to second order in the $q$-fields,
\begin{align}
\begin{split}
    a^{-3}\cl^{(0)} &=3H^2(2-\epsilon)-3\xi H^4(1-\epsilon)+\tfrac{1}{2}(\dot\chi^2+\dot\phi^2+\phi^2\dot\theta_c^2)-V~,
\end{split}\\
\begin{split}
    a^{-3}\cl^{(1)} &=\big[3H^2(1-\omega)-\tfrac{1}{2}(\dot\chi^2+\dot\phi^2+\phi^2\dot\theta_c^2)-V\big]\alpha+\phi^2\dot\theta_c\dot\theta_q~,
\end{split}\\
\begin{split}
    a^{-3}\cl^{(2)} &=-\big[3H^2(1-2\omega)-\tfrac{1}{2}(\dot\chi^2+\dot\phi^2+\phi^2\dot\theta_c^2)\big]\alpha^2-(2-3\omega)H\alpha\frac{\partial_i^2}{a^2}\beta\\
    &\hspace{13pt} +\tfrac{1}{2}\phi^2\Big(\dot\theta_q^2-2\dot\theta_c\alpha\dot\theta_q+\theta_q\frac{\partial_i^2}{a^2}\theta_q+2\dot\theta_c\theta_q\frac{\partial_i^2}{a^2}\beta\Big)~,
\end{split}
\end{align}
where $V=\cu(\chi)+\tfrac{\lambda}{4}(f^2-\phi^2)^2$~.

In linear perturbation theory, background fields and perturbations decouple from each other. By varying $\cl^{(0)}$ one obtains the second Friendmann equation as well as background equations for relevant scalar fields, including the inflaton. The first Friedmann equation is obtained by varying $\cl^{(1)}$ w.r.t. $\alpha$. From $\cl^{(2)}$ one can then obtain equations for the relevant perturbations. In stochastic formalism, perturbations and background fields mix due to time-varying window function introduced earlier. In our scenario, we assume that inflation is primarily driven by the inflaton $\chi$, and the effects of the PQ fields $\phi$ and $\theta$ on inflation can be ignored. This assumption simplifies the resulting equations of motion. For example, the coefficient of $\alpha$ in $\cl^{(1)}$ vanishes, as $3H^2\simeq V_{\rm inf}$ while the other terms are much smaller in comparison. This means that the equations for $\alpha$ and $\beta$ are the same as in the linear perturbation theory and coincide with \eqref{alpha_eq} and \eqref{beta_eq} (after replacing $\delta\theta$ with $\theta_q$). When deriving the EOM for $\theta_q$ (or $\theta_c$), the last term in $\cl^{(1)}$ mixes $c$- and $q$-fields, and by varying $\cl^{(1)}+\cl^{(2)}$ w.r.t. $\theta_q$, we get (after eliminating $\alpha$ and $\beta$)
\begin{equation}\label{stochastic_eq_t}
    \ddot\theta_c+\Big(3H+2\frac{\dot\phi}{\phi}\Big)\dot\theta_c\simeq -\ddot\theta_q-\Big(3H+2\frac{\dot\phi}{\phi}\Big)\dot\theta_q+\ldots~,
\end{equation}
where we used the slow-roll conditions \eqref{SR_epsilon_omega}, and the ellipsis contains terms without time-derivatives of $\theta_q$. Such terms are irrelevant, since they will be cancelled after using the windowed Fourier transoform \eqref{Fourier_theta_q} and the equation of motion for $\delta\theta_k$, which is given by \eqref{d_theta_k_eq} under the conditions
\begin{equation}\label{dphi_conditions}
    \Big|\frac{\dot\phi}{H\phi}\Big|,\Big|\frac{\ddot\phi}{H^2\phi}\Big|\ll 1~.
\end{equation}

Let us now re-write \eqref{stochastic_eq_t} in the e-fold time, assume the condition \eqref{dphi_conditions}, and use the Fourier transform \eqref{Fourier_theta_q}. This results in stochastic (massless) Klein--Gordon equation
\begin{equation}\label{stochastic_KG}
    \theta_{c,NN}+3\theta_{c,N}\simeq\Xi_{\theta,N}+\Xi_{\Pi}+3\Xi_{\theta}~,
\end{equation}
where the right side consists of the noise terms
\begin{align}
    \Xi_{\theta} &\equiv -\int\frac{d^3k}{(2\pi)^{3/2}}W_{,N}\Big(\frac{k}{k_\sigma}\Big)(e^{-i{\bf kx}}\delta\theta_k \hat a_k+{\rm h.c.})~,\label{theta_noise}\\
    \Xi_{\Pi} &\equiv -\int\frac{d^3k}{(2\pi)^{3/2}}W_{,N}\Big(\frac{k}{k_\sigma}\Big)(e^{-i{\bf kx}}\delta\theta_{k,N} \hat a_k+{\rm h.c.})~\label{Pi_noise}.
\end{align}
We introduced in advance the symbol $\Pi$ for canonical momentum (in e-fold time) of the ALP, $\Pi=\theta_{,N}$, and $\Xi_{\theta}$ and $\Xi_{\Pi}$ are the field and momentum noises. The momentum is also split as $\Pi=\Pi_c+\Pi_q$ with
\begin{equation}
    \Pi_q=\int\frac{d^3k}{(2\pi)^{3/2}}W\Big(\frac{k}{k_\sigma}\Big)(e^{-i{\bf kx}}\delta\theta_{k,N} \hat a_k+{\rm h.c.})~.
\end{equation}
By combining the definition of the momentum with \eqref{stochastic_KG}, we finally obtain the Langevin equations,
\begin{align}
\begin{split}\label{Langevin_1}
    \theta_{c,N} &\simeq \Pi_c+\Xi_\theta~,
\end{split}\\
\begin{split}\label{Langevin_2}
    \Pi_{c,N} &\simeq -3\Pi_c+\Xi_\Pi~,
\end{split}
\end{align}
where the correlation functions of $\Xi_\theta$ and $\Xi_\Pi$ can be derived from \eqref{theta_noise} and \eqref{Pi_noise}. Under the conditions \eqref{SR_epsilon_omega} and \eqref{dphi_conditions}, we have
\begin{equation}
    \langle\Xi_{\theta}(N)\Xi_{\theta}(N')\rangle\simeq \frac{k^3_{\sigma}}{2\pi^2}|\delta\theta_{k_\sigma}|^2\delta(N-N')\simeq\Big(\frac{H}{2\pi\phi}\Big)^2\delta(N-N')~,\label{Xi_two_point_fn}
\end{equation}
while $\langle\Xi_{\theta}(N)\Xi_{\Pi}(N')\rangle$ and $\langle\Xi_{\Pi}(N)\Xi_{\Pi}(N')\rangle$ are suppressed by the slow-roll parameters ($\epsilon$, $\omega$, and their rates of change).

Ignoring $\Xi_\Pi$ in \eqref{Langevin_2}, we can solve it as $\Pi_c\simeq\Pi_c(N_{\rm in})e^{3(N_{\rm in}-N)}$, and plug it in \eqref{Langevin_1} which results in
\begin{equation}\label{theta_USR_noise}
    \theta_{c,N}\simeq\theta_{c,N}(N_{\rm in})e^{3(N_{\rm in}-N)}+\Xi_\theta~.
\end{equation}
In the absence of noise, \eqref{theta_USR_noise} coincides (after integration) with the solution \eqref{theta_USR_sol}.

In summary, standard results for the ALP fluctuations in (quasi-) de Sitter background can be reproduced under the conditions \eqref{SR_epsilon_omega}, \eqref{theta_N_condition}, and \eqref{dphi_conditions}. Then one can follow the standard Fokker--Planck equation, since it can be derived from these Langevin equations.

\subsection{Note on initial condition for $\phi$ after the symmetry breaking}

During the first, symmetric, phase of inflation, $\phi$ is stabilized at $\phi=0$ by its effective time-dependent mass. This effective mass vanishes at some time $N_c$ (before turning tachyonic), after which symmetry breaking is triggered by quantum fluctuations of $\phi$. Thus, the classical evolution of $\phi$ can be divided into the symmetric phase and the broken phase. Between the two classical phases we have a period of quantum diffusion domination (described by stochastic formalism) when the effective mass of $\phi$, and its velocity are nearly zero. In order to solve the classical background equations for the post-quantum-diffusion phase, the initial conditions for $\phi$ must be set, which can be estimated from the results of stochastic formalism describing the quantum diffusion. Unlike in the case of the ALP, the quantum diffusion phase of $\phi$ does not last until the end of inflation: as the tachyonic mass of $\phi$ grows with time, the classical velocity of $\phi$ grows as well, and soon dominates over the quantum fluctuations. The precise time when the evolution of $\phi$ classicalizes depends on the parameter choice, namely on the magnitude of the effective tachyonic mass: the faster the (tachyonic) mass grows, the faster the evolution of $\phi$ classicalizes. Similar treatment of a quantum diffusion phase can be found in hybrid inflation scenarios during waterfall transition, see for example Refs. \cite{Garcia-Bellido:1996mdl,Clesse:2010iz,Clesse:2015wea,Kawasaki:2015ppx,Tada:2023pue,Aldabergenov:2024fws}.

The stochastic (slow-roll) evolution of $\phi$ is described by \cite{Aldabergenov:2025ulq}
\begin{equation}\label{Langevin_phi}
    \phi_{,N}\simeq -\frac{V_{\rm eff,\phi}}{3H^2}+\Xi_{\phi}\simeq -\frac{m^2_{\rm eff}}{3H^2}\phi+\Xi_{\phi}~,
\end{equation}
where we linearized the potential term around the symmetric point $\phi=0$, and
\begin{equation}
    m^2_{\rm eff}=V_{\rm eff,\phi\phi}=\alpha H^4-\lambda f^2~.
\end{equation}
The noise $\Xi_\phi$ satisfies
\begin{equation}
    \langle\Xi_{\phi}(N)\Xi_{\phi}(N')\rangle\simeq\Big(\frac{H}{2\pi}\Big)^2\delta(N-N')~.\label{Xi_phi_two_point_fn}
\end{equation}

From \eqref{Langevin_phi} one can obtain the equation for the evolution of the quadratic expectation value of $\phi$,
\begin{equation}\label{phi^2_stochastic}
    \langle\phi^2\rangle_{,N}\simeq -\frac{2m^2_{\rm eff}}{3H^2}\langle\phi^2\rangle+ \Big(\frac{H}{2\pi}\Big)^2~.
\end{equation}
Around the critical point $N_c$ one can further simplify the equation by linearizing in $N_c-N$. For this, we need the analytical expression for the evolution of the inflaton $\chi$ (in order to estimate how $m_{\rm eff}^2$ changes with $N$). For effectively single-field inflation the inflaton evolution is given by \cite{Aldabergenov_2025}
\begin{equation}
    y\equiv e^{-\chi}\simeq\frac{y_*}{1-2y_*N}~,
\end{equation}
where $y_*\equiv e^{-\chi_*}$ is the inflaton value at the CMB reference scale, and $N=0$ at the start of observable inflation. Expanding $y(N)$ around the critical time, we get
\begin{equation}
    y\simeq y_c-2y_c^2(N_c-N)~,
\end{equation}
where $y_c$ is the inflaton value at the critical time. With the above result we can write
\begin{equation}
    H^2\simeq \tfrac{1}{3}\cu\simeq\tfrac{1}{6}M^2(1-y_c)\big[1-y_c+4y_c^2(N_c-N)\big]~,
\end{equation}
so that
\begin{equation}
    m^2_{\rm eff}(N)\simeq \mu(N_c-N)~,~~~\mu\equiv\frac{8\lambda f^2y_c^2}{1-y_c}~,\label{m_eff_linearlized}
\end{equation}
where we used the fact that (by definition of the symmetry breaking time)
\begin{equation}
    m^2_{\rm eff}(N_c)=\frac{\alpha}{36}M^2(1-y_c)^4-\lambda f^2=0~,
\end{equation}
and $\alpha M^2$ can be expressed in terms of $\lambda f^2$.

\begin{figure}[H]
	\begin{center}
\includegraphics[width=0.8\textwidth]{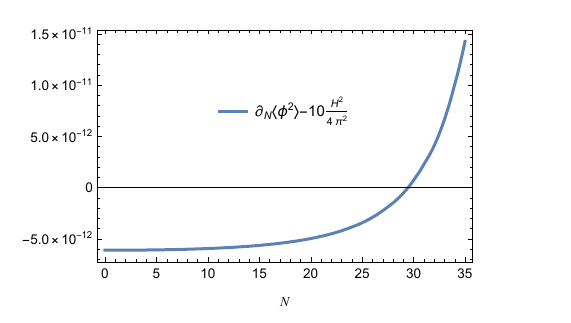}
	\end{center}
\caption{Numerical solution of \eqref{phi^2_stochastic}.  Here we utilize condition \eqref{Classicality_condition} to determine the e-fold at which $U(1)$ symmetry is broken, i.e. field $\phi$ becomes classic and rolls down to the potential's minimum.}
	\label{efoldtoclassic}
\end{figure}

By using \eqref{m_eff_linearlized}, Eq. \eqref{phi^2_stochastic} can be solved as
\begin{equation}\label{phi_erf}
    \langle\phi^2\rangle\simeq \frac{M^3}{48\pi^2}\sqrt{\frac{\pi}{2\mu}}\exp\Big[\frac{2\mu}{M^2}(N_c-N)^2\Big]\Big\{1-{\rm erf}\Big[\frac{\sqrt{2\mu}}{M}(N_c-N)\Big]\Big\}~,
\end{equation}
where the integration constant is chosen such that $\langle\phi^2\rangle\rightarrow 0$ as $N\rightarrow -\infty$, and $y_{\rm PQ}\ll 1$ was used (in our models of interest, symmetry breaking occurs sufficiently early during inflation when $y$ is still small). To obtain an initial condition for $\phi$ and its velocity for the subsequent classical phase, we use Eq. \eqref{phi_erf} and write
\begin{equation}\label{initial_conditions_phi}
    \phi(N_{\rm cl})=\sqrt{\langle\phi^2\rangle}\Big|_{N=N_{\rm cl}}~,~~~\phi_{,N}(N_{\rm cl})=\partial_N\sqrt{\langle\phi^2\rangle}\Big|_{N=N_{\rm cl}}~,
\end{equation}
where $N_{\rm cl}$ is the time where the evolution of $\phi$ classicalizes, i.e., stochastic noise contribution $H^2/(4\pi^2)$ becomes subdominant to the velocity term $\partial_{,N}\langle\phi^2\rangle$. This point in time ($N_{\rm cl}$) can be used as the initial time for the next classical phase. For example we can use
\begin{equation}\label{Classicality_condition}  \partial_{,N}\langle\phi^2\rangle=10\times  H^2/(4\pi^2)~,
\end{equation}
as an equation to find $N_{\rm cl}$. The $N_{cl}$ found via condition \eqref{Classicality_condition} changes negligibly if one assume another coefficient instead of $10$ before the noise, e.g. $15$, which makes this condition robust, though it still can be considered heuristic. We present numerical solution in Fig.~\eqref{efoldtoclassic}. The higher the numerical coefficient in front of $H^2/(4\pi^2)$ is, the less important the quantum diffusion effects are (this also pushes $N_{\rm cl}$ farther away to an extent). Since during quantum diffusion $\phi$ and $\phi_{,N}$ fluctuate around zero, the phase with broken symmetry (when the ALP is generated) can be defined once the classical evolution of $\phi$ (away from zero) takes over.

We note that the solution \eqref{phi_erf} is accurate in the vicinity of the critical time while $|N-N_c|\ll 1$, so if the quantum phase lasts more than one e-fold (starting from $N_c$), one should instead solve Eq. \eqref{phi^2_stochastic} numerically.

\begin{figure}[H]
	\begin{center}
\includegraphics[width=0.8\textwidth]{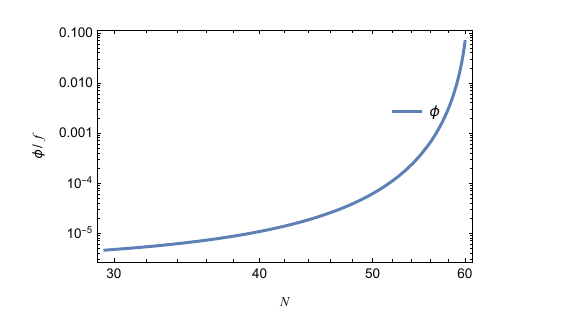}
	\end{center}
\caption{Numerical solution to the equation of motion for radial part of the PQ field $\phi$ with utilization of initial conditions defined in \eqref{initial_conditions_phi}.}
	\label{PhiRadial}
\end{figure}

Now given with initial condition for field $\phi$, we present numerical solution to its classical equation of motion in Fig.~\eqref{PhiRadial}. 

\begin{figure}[H]
	\begin{center}
\includegraphics[width=0.8\textwidth]{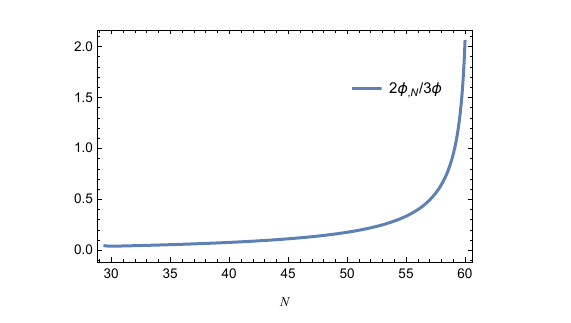}
	\end{center}
\caption{Slow-roll parameter for $\phi$.}
	\label{PhiSlow}
\end{figure}

Now we should check the slow-roll condition for $\phi$. Given the numerical solution in Fig.~\eqref{PhiRadial}, we present slow-roll parameter in Fig.~\eqref{PhiSlow}.

\begin{figure}[H]
	\begin{center}
\includegraphics[width=0.8\textwidth]{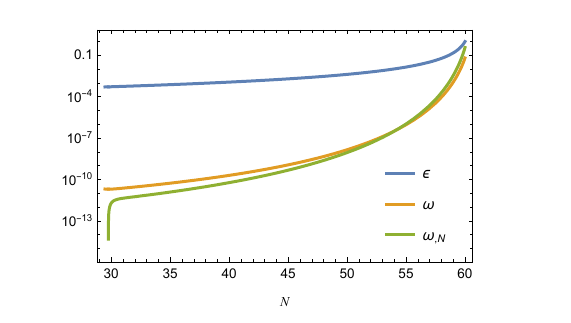}
	\end{center}
\caption{Inflaton ($\epsilon$) and GB ($\omega,\, \omega_N$) slow-roll parameters. This plot indicates the validity of our single-field inflation approximation.}
	\label{slowrolling}
\end{figure}

Next we would like to check the condition \eqref{no_backreaction_condition}. We show the numerical solutions in Fig.~\eqref{slowrolling}.

\section{Power spectrum of ALP}

\begin{figure}[H]
	\begin{center}
\includegraphics[width=0.7\textwidth]{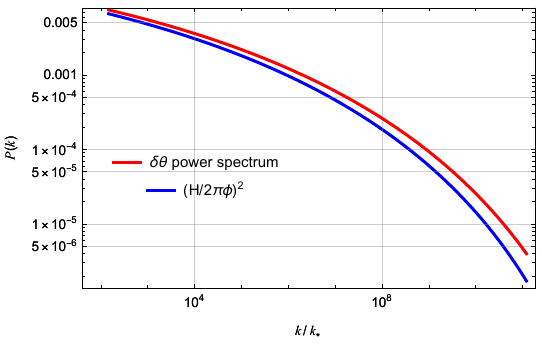}
	\end{center}
\caption{ALP power spectrum calculated with assumption $a(N)=e^{N}$ and $k_*=He^{N_{cl}}$. The red line is obtained via numerical solution to Eq.~\eqref{u_k_eq_N}. The blue line is semi-analytical approximation under our assumptions. Power spectrum is found to be strongly scale-dependent and at largest scales available values difference is about $10\%$.}
	\label{PhaseFlucts}
\end{figure}

The ALP power spectrum is
\begin{equation}
    \cp_{\delta\theta}(k)=\cfrac{k^3}{2\pi^2}\left| \delta\theta_k\right|^2~,
\end{equation}
calculated at the horizon exit. The approximated solution for $\delta\theta_k$ is given by \eqref{dtheta_SR_solution}, so that the power spectrum reads
\begin{equation}
    \cp_{\delta\theta}(k)\simeq \frac{H^2}{4\pi^2\phi^2}\Big|_{aH=k}~,
\end{equation}
and is shown in Fig.~\eqref{PhaseFlucts}.

We want to relate our $k$-modes of interest to the CMB reference scale, such as $k_*=0.05~{\rm Mpc^{-1}}$. This can be done by introducing ``physical" wavenumber
\begin{equation}\label{PhysK}
    k_{\rm phys}=\frac{k_i}{k_0}\times 0.05~({\rm Mpc^{-1}})~,
\end{equation}
where $k_i$ is our chosen set of modes, and $k_0$ is the mode that leaves the horizon at $N=0$, i.e. the CMB mode. When $k_i=k_0$ we get a normalized CMB mode $k_{\rm phys}=0.05~{\rm Mpc^{-1}}$, and all other subsequent modes will be normalized accordingly. For our ALP case, the modes that leave the horizon before $N_{\rm cl}$ are not well-defined because the ALP does not exist yet in the symmetric phase, but $k_0$ should still be defined from $k_0=a(0)H(0)$, in order to relate our modes to the CMB reference scale.

From Fig.~\ref{PhaseFlucts}, where $a(N)=e^N$ and $k_*=e^{N_{cl}}H$ we can estimate the scale corresponding to the peak of the power spectrum. 
Now, using Eq.~\eqref{PhysK}:
\begin{equation}
    k_{phys}\approx \cfrac{e^{N_{cl}}H}{H} \times 0.05 (\text{Mpc}^{-1})\approx 5.3\cdot 10^{11}\text{Mpc}^{-1}\rightarrow k_{phys}^{-1}\approx 1.87\cdot 10^{-12}\text{Mpc}\sim 0.4 \,Au,
\end{equation}
which corresponds to the scales smaller than Earth's orbit. This scale corresponds to the maximal value of power spectrum.

In the Langevin equation \eqref{theta_USR_noise} for the ALP, we can ignore the initial velocity (as justified earlier), and write
\begin{equation}\label{axion_Langevin}
    \theta_{c,N}\simeq\Xi_\theta~,
\end{equation}
where the noise satisfies \eqref{Xi_two_point_fn}, or written in terms of the power spectrum,
\begin{equation}
    \langle\Xi_{\theta}(N)\Xi_{\theta}(N')\rangle\simeq \cp_{\delta\theta}\,\delta(N-N')~.\label{Xi_two_point_fn_P}
\end{equation}
Then the FP equations for the probability distribution $P(\theta,N)$ is as follows \cite{Noorbala_2024}:
\begin{equation}
    \cfrac{\partial P(\theta,N)}{\partial N}=\frac{1}{2}\cp_{\delta\theta}\cfrac{\partial ^2 P(\theta,N )}{\partial \theta^2}~,
\end{equation}
where we dropped subscript $c$ for simplicity, so that $\theta$ should be understood as the effective background value of the ALP. Initial condition is $P(\theta,N_{PQ})=\delta(\theta-\theta_i)$.

Solution is then given by:
\begin{equation}
    P(\theta,t)=\cfrac{1}{\sqrt{2\pi \Sigma(t)}}\exp{\left( -\cfrac{(\theta-\theta_i)^2}{2\Sigma(t)}\right)},
\end{equation}
where 
\begin{equation}
    \Sigma(N)=\int_{N_{PQ}}^N \cp_{\delta\theta}(N')dN'\approx \int_{N_{PQ}}^N \left( \cfrac{H}{2\pi {\phi}}\right)^2dN'.
\end{equation}
The probability of crossing $\pi$ in e-fold $N$ is given by:
\begin{equation}\label{cross_prob}
    p(N)=\int_\pi^\infty P(\theta,N)d\theta = \cfrac{1}{2}\,\text{erfc}\left(\cfrac{\pi-\theta_i}{\sqrt{2}\Sigma(N)} \right),
\end{equation}
which demonstrates strong dependence on initial phase. 
Number of such crossings:
\begin{equation}
    n_c=p(N)e^{3N}.
\end{equation}
With the parameters chosen, the number of crossing is $n_c\approx 2.28\cdot 10^{14}$, which corresponds to the number of possible domain walls.

\section{Potential cosmological scenarios}
In this section we focus on the last unfixed parameter~--- $\Lambda$, which defines the energy scale of ALP's potential. Let us investigate possible cosmological scenarious.

\subsection{Non-relativictic ALP matter}
Given with parameters fixed in previous sections and assuming ALP is representing at least a part of dark matter, we will calculate the value of $\Lambda$.

ALP's energy density (misalignment ALP) is calculated as:
\begin{equation}
    \rho_{harm}=\cfrac{1}{2}m_a^2f^2\theta_i^2.
\end{equation}
There are also ways to take into account anharmonicity \cite{Visinelli_2009, Bae_2008}:
\begin{equation}
    \rho_{anharm}=\cfrac{1}{2}m_a^2f^2\theta_i^2F(\theta_i),
\end{equation}
where 
\begin{equation}    F(\theta_i)=\left[\ln{\left( \cfrac{e}{1-\theta_i^2/\pi^2}\right)} \right]^{7/6}.
\end{equation}
Let us estimate average square of the phase assuming uniform distribution and compare it with the $\Sigma$:
\begin{equation}
  <\theta^2_i>=\cfrac{1}{2\pi}\int_{-\pi}^\pi \theta^2d\theta=\cfrac{\pi^2}{3}\gg \Sigma(60),
\end{equation}
which demonstrates we can ignore the impact of quantum fluctuations on random distribution of the phase.

In our case we have 
\begin{equation}
    \rho_{ALP,i}=\cfrac{1}{2}\Lambda^4<\theta^2_i>F(\sqrt{<\theta^2_i>})\approx 2.5\Lambda^4.
\end{equation}

Let us consider simple condition for oscillations start:
\begin{equation}
 3H=m_{ALP}=\Lambda^2/f, 
\end{equation}
which implies the temperature at the start of the oscillations $T_{osc}$ is found as follows (in reduced Planck units):
\begin{equation}
    3\cdot 1.66\sqrt{g_*(T_{osc})}T^2_{osc}\approx \Lambda^2/f,
\end{equation}
which implies:
\begin{equation}
    T_{osc}\approx \cfrac{\Lambda}{\sqrt{5f\sqrt{g_*(T_{osc})}}}.
\end{equation}
Modern ALP density could be estimated as follows:
\begin{equation}\label{ALP_dens}
    \rho_{ALP,0}=\rho_{ALP,i}\cdot \left(\cfrac{ a_{osc}}{a_{0}}\right)^3=\rho_{ALP,i}\cdot \left(\cfrac{ T_0}{T_{osc}}\right)^3\approx 28\Lambda (g_*(T_{osc}))^{3/4}f^{3/2}T_0^3,
\end{equation}
where modern temperature $T_0\approx 2.6\cdot 10^{-13} GeV$.

From expression \eqref{ALP_dens}:
\begin{equation}
    \rho_{ALP,0}\approx2.8\cdot 10^{-21}\cdot\cfrac{\Lambda (GeV)}{10^{18}GeV}GeV^4.
\end{equation}
Taking $\Lambda = 10^{-10}GeV$ we obtain:
\begin{equation}
    \rho_{ALP,0}\approx2.8\cdot 10^{-49}GeV^4
\end{equation}

Modern critical density is\begin{equation}
    \rho_0=\cfrac{3H_0^2}{8\pi G}\approx 8\cdot 10^{-47}GeV^4.
\end{equation}
With such a value of $\Lambda$, the mass of the particle $m_{ALP}=\cfrac{\Lambda^2}{f} \leq 10^{-29}eV$ defines it as ultra-light boson, which links our model to fuzzy dark matter scenarious. For extensive review, please, see \cite{eberhardt2025ultralightfuzzydarkmatter} and references therein. At the current moment constraints are imposed on fuzzy dark matter in a wide range of boson's masses \cite{2024PhRvD.110l3532L}. 

\subsection{ALP Dark energy scenario}
Modern Hubble parameter is equal to
\begin{equation}
    H_0\sim 10^{-42}GeV
\end{equation}
and dark energy constitutes $\Omega_{DE}\approx 0.685$ \cite{2020} of the critical density of the Universe. Given with ALP energy density of order $\Lambda^4$, we can estimate the value of $\Lambda$ to make our ALP a dark energy candidate:
\begin{equation}
    \Lambda^4=\Omega_{DE}\rho_0 \rightarrow \Lambda\approx 2.7\cdot10^{-12}GeV,
\end{equation}
then the mass of the ALP is $m_{ALP}\approx 1.9\cdot 10^{-42}GeV$, which implies particle's mass small enough for its classical motion to be effectively frozen, since $3H_0 > m_{ALP}$. 

Given with power spectrum peak at scales near astronomical unit, this scenario would provide inhomogeneous dark energy, which eventually would become non-relativistic matter. This refers to the local-to-global evolution of the universe. This field would affect the cosmological evolution only at modern times, since its energy density is small. Such non-homogeneous structure of dark energy and its dynamical nature could provide a solution to the Hubble tension problem \cite{Riess_2022, 2020}.

We also would like to note the following feature of this particular model. Due to the self interaction of the ALP field, there would be a smooth transition of phase between different regions. Such smooth transition will eventually become closed forming string. This will lead to the formation of the network of domain walls with strings. Such structure, if dominate the Universe, would provide equation of state parameter $\omega$ in the range $-2/3\leq \omega \leq -1/3$, providing accelerated expansion. Such structures are studied in, e.g. \cite{1990ApJ...357..293R, Hiramatsu_2011}, they are found to be unstable.

\subsubsection{Resolving the Hubble tension?}
Neglecting radiation and assuming dust-matter is distributed uniformly across the Universe, let us define
\begin{equation}
    \sigma_H^2=\langle H^2_i\rangle-\langle H_i\rangle^2,
\end{equation}
where subscript $i$ refers to regions with different values of initial phase. 
Following the approach demonstrated in \cite{Raffai:2025ldr}, which refers to local-to-global evolution of the universe, we can estimate the deviation:
\begin{equation}
    \sigma^2_{H}=\cfrac{\langle H^2_i\rangle}{4}\langle\widetilde{\Omega}^2_{ALP,i} \rangle,
\end{equation}
where 
\begin{equation}
    \widetilde{\Omega}_{ALP,i}=\Omega_{ALP,i}-\langle\Omega_{ALP,i} \rangle,
\end{equation}
where $\langle\Omega_{ALP,i} \rangle$ is defined as follows:
\begin{equation}
    \langle\Omega_{ALP,i} \rangle=\cfrac{1}{2\pi}\int^\pi_{-\pi} \cfrac{\Lambda^4(1-\cos{\theta})}{\rho_m+\Lambda^4(1-\cos{\theta})}d\theta.
\end{equation}
Numerical integration yields 
$\sigma_H\approx 0.13\sqrt{\langle H^2_i\rangle}$,
which allows to solve Hubble tension on small scales.


\subsection{Condition for black hole formation}
Let us now study the possibility of black hole formation in our model. From
\eqref{cross_prob} with $\theta_i=\sqrt{<\theta_i^2>}$ we have strong suppression of crossings over potential's maximum. Therefore probability of domain wall formation is strongly suppressed.

Let us now estimate maximum and minimal masses of probable black holes.

Assume that these fluctuations would enter Hubble horizon during RD, then size of the fluctuations after entering horizon are as follows:
\begin{equation}  r(N)=\cfrac{e^{2(N_{inf}-N)}}{2HN_{inf}}.
\end{equation}
Size of the fluctuation allows to estimate its mass:
\begin{equation}\label{r_re-entry}
    r(N)=\sqrt{\cfrac{m}{4\pi \sigma}}=\cfrac{e^{2(N_{inf}-N)}}{2HN_{inf}}.
\end{equation}

After horizon re-entry a closed domain wall typically starts to shrink until its radius reaches its thickness $d=f/(2\Lambda^2)$. Therefore, to form a black hole it needs to satisfy $d<r_s$, where $r_s\equiv m/(2\pi)$ is its Schwarzschild radius (in reduced Planck units). By using \eqref{r_re-entry} we can write the mass of the wall (upon horizon re-entry) as
\begin{equation}
    m(N)=\frac{4\pi\Lambda^2f}{H_{inf}^2N_{inf}^2}e^{N_{inf}-N}~.
\end{equation}
Here $N$ corresponds to the forward e-fold number during inflation (taking $N=0$ at the start of observable inflation, and $N=N_{inf}$ at the end), at which the corresponding ALP fluctuation is generated. Hence, from the collapse condition $d<r_s$ we obtain
\begin{equation}
    N<N_{inf}-\frac{1}{2}\ln\Big(\frac{H_{inf}N_{inf}}{\sqrt{2}\Lambda^2}\Big)\equiv N_{max}~,
\end{equation}
implying that only the fluctuations (domain wall seeds) generated before $N_{max}$ eventually collapse into black holes. The walls whose seeds are produced after $N_{max}$ have their thickness exceeding their Schwarzschild radius.

Let us now estimate maximal mass:
\begin{equation}
    m_{max}=16\pi\Lambda^2 f\left( \cfrac{e^{2(N_{inf}-N_{cl})}}{2HN_{inf}}\right)^2\approx 2.3\cdot 10^{-15}M_{\odot}.
\end{equation}

Now estimate minimal mass from the condition width of the wall should be less than its Schwarzschild radius:
\begin{equation}
  m_{min}>\cfrac{ f}{4 G\Lambda^2} \approx 0.5 M_{\odot}.
\end{equation}

In both estimations we have adopted $\Lambda=2.7\cdot10^{-12}GeV$. Since $m_{min}\gg m_{max}$ there is no way to form a black hole in our particular scenario.

\section{Constraints on parameters of the ALP}
Let us now discuss possible ways to constraint our model.

\begin{itemize}
    \item The trans-planckian censorship constraints  (TCC) paper \cite{SHLIVKO2023138251} provides upper limit on the value of $f$ (for $m_{ALP}<1GeV$):
\begin{equation}
    f\leq 0.6 M_{Pl}.
\end{equation}
Since the TCC prohibits this for trans-Planckian modes, one gets an upper bound on the duration of such accelerated phases so that a Planck length mode never exits the Hubble horizon. For inflation, this implies \cite{Bedroya_2020}
\begin{equation}
    e^N<M_{Pl}/H_{inf},
\end{equation}
where N is the number of e-folds of inflation and we assume that the Hubble parameter $H_{inf}$ remains constant during inflation.
This TCC concept is not widely adopted and we refer to it just for review.

We do not satisfy this particular constraint in our calculations.

\item Isocurvature constraints.

In the paper \cite{Kawasaki_2018} there is much stronger upper bound for $f$:
\begin{equation}
    f\sim 10^{12}GeV,
\end{equation}
however, it follows from isocurvature constraints on CMB scale, which is not the case in our scenario, since in our model ALP is not defined at the start of observable inflation.

\item Superradiance constraints from merger of two $\sim 100M_{\odot}$ BHs \cite{caputo2025superradianceconstraintsgw231123}:
\begin{equation}
    f\geq 10^{14}GeV,\, m_{ALP}\sim 10^{-13}eV.
\end{equation}
Superradiance is an effect of enhancement of radiational emission \cite{Brito_2020}. The effect is as follows: light bosons take energy away from BH, slowing down its rotation. Constraint on specific mass of the light bosons (e.g. ALPs) are following from the observation of rapidly rotating BHs in a certain mass range.  The effect is most relevant if mass of the boson $\mu$ is about \cite{witte2025steppingsuperradianceconstraintsaxions}:
\begin{equation}
    \mu\sim \cfrac{0.1}{GM_{BH}},
\end{equation}
which for our parameters $m_{ALP}=\cfrac{\Lambda^2}{f} \leq 10^{-29}eV$ provides
\begin{equation}
    M_{BH}\sim 10^{19}M_{\odot}.
\end{equation}
There is no data in favor of existence of such a gigantic black holes. Therefore, no constraints.

\item Since our ALP can possess a mass $m_{ALP}\sim10^{-29}eV,$ which makes it a fuzzy dark matter candidate, it is a subject to constraints on fuzzy dark matter \cite{2024PhRvD.110l3532L}. Given with other parameters fixed, it constraints the value of $\Lambda\leq 10^{-10}$ GeV.

However, in case of dark energy scenario, this constraint is not applicable.


\end{itemize}

\section{Stochastic gravitational waves from closed wall collapse}
\label{sec:gw}

The same domain wall sector that can seed the primordial black-hole population also provides a possible source of a stochastic gravitational wave background. This is a natural consequence of the fact that closed walls do not remain static: after formation they collapse, and any departure from exact spherical symmetry allows part of the released energy to be emitted in gravitational waves. Gravitational radiation from collapsing vacuum domains and cosmic domain walls has been studied in several related contexts \cite{Gleiser:1998gw,Hiramatsu_2010,Hiramatsu_2014,Saikawa_2017,Ferreira_2023}, and the corresponding nanohertz signal is phenomenologically relevant in view of the NANOGrav evidence for a stochastic background \cite{Agazie_2023_GWB,Agazie_2023_Detector}. In the present model this part should be understood as a late time phenomenological extension of the domain wall sector.

The dimensional scale is the same fixed one used in the previous sections. Thus the gravitational-wave calculation is performed with the core parameters
\begin{equation}
\lambda_{\rm PQ}=6.3096\times 10^{-11},
\quad
A=\frac{f}{M}=1.25\times 10^{5},
\quad
b=37.25,
\quad
f=1.6272\times 10^{16}\ {\rm GeV},
\end{equation}
together with
\begin{equation}
\Lambda =  10^{-10}\ {\rm GeV}.
\end{equation}
Here \(\lambda_{\rm PQ}\) is the dimensionless coupling, while \(\Lambda\) is the dimensional scale entering the wall potential. The remaining derived quantities are
\begin{equation}
M=1.3018\times 10^{11}\ {\rm GeV},
\qquad
\alpha_{\rm GB}=2.1982\times 10^{-21}\ {\rm GeV}^{-2}.
\end{equation}
Therefore the gravitational wave calculation does not change the inflationary normalization, the Gauss--Bonnet coupling, or the fixed wall-potential scale.

The wall potential is written as
\begin{equation}
V_{\rm wall}(\theta)
=
\Lambda^{4}
\left(1-\cos\theta\right),
\end{equation}
so that the wall tension is
\begin{equation}
\sigma_{\rm wall}
=
4f\Lambda^{2}.
\end{equation}
With the fixed values above this gives
\begin{equation}
\sigma_{\rm wall}
=
6.51\times 10^{-4}\ {\rm GeV}^{3}.
\end{equation}

The closed wall mass distribution is modeled by a log normal window,
\begin{equation}
W(M)
=
\frac{1}{\sqrt{2\pi}\sigma_{\ln M}}
\exp\left[
-\frac{\left(\ln M-\ln M_c\right)^2}{2\sigma_{\ln M}^{2}}
\right],
\end{equation}
normalized as
\begin{equation}
\int W(M)\,d\ln M=1.
\end{equation}
The corresponding differential wall abundance is
\begin{equation}
\frac{d\beta_{\rm wall}}{d\ln M}
=
\beta_{\rm wall}^{\rm tot}W(M).
\end{equation}
The probability of producing closed walls is controlled by the displacement of the angular field from the maximum of the periodic potential. We use
\begin{equation}
P_{\rm wall}
=
\frac{1}{2}
{\rm erfc}
\left[
\frac{\pi-\theta_i}{\sqrt{2}\sigma_\theta}
\right].
\end{equation}
The scan varies the late wall-production and collapse shape sector, namely the closed wall abundance, the mass window parameters, the stochastic angular width, and the nonsphericity parameters. These quantities affect the amplitude and shape of the gravitational-wave spectrum.

The energy fraction emitted into gravitational waves is described phenomenologically as
\begin{equation}
\epsilon_{\rm GW}(M,e)
=
\epsilon_{{\rm GW},0}\,
e^{2}\,
{\cal C}(M,e),
\end{equation}
where \(e\) parameterizes the nonsphericity of the collapsing wall and \({\cal C}(M,e)\) is the compactness-like collapse factor. Perfectly spherical collapse does not efficiently radiate, so the factor \(e^2\) captures the leading dependence on the quadrupolar deformation. The eccentricity distribution is restricted by
\begin{equation}
0\leq e\leq e_{\rm max}.
\end{equation}
Thus the detectable gravitational wave branch requires sufficiently nonspherical and sufficiently compact wall collapse.

The present day peak frequency associated with a wall of mass \(M\) is estimated as
\begin{equation}
f_0(M,e)
=
\frac{a_{\rm col}}{a_0}
\frac{{\cal C}(M,e)}{2\pi R_{\rm col}(M,e)}.
\end{equation}
Using entropy conservation,
\begin{equation}
\frac{a_{\rm col}}{a_0}
=
\frac{T_0}{T_{\rm col}}
\left(
\frac{g_{s,0}}{g_{s,{\rm col}}}
\right)^{1/3}.
\end{equation}
The spectral shape is modeled by a broken power law,
\begin{equation}
{\cal S}(x)
=
\frac{1+q}{q x^{-3}+x^{q}},
\qquad
x=\frac{f}{f_0}.
\end{equation}
This form has the expected causal low frequency behavior, \({\cal S}\propto f^3\), and a phenomenological high frequency tail controlled by \(q\). The present day stochastic background is then
\begin{equation}
\Omega_{{\rm GW},0}(f)
=
\Omega_{r,0}
\left(
\frac{g_{*,{\rm col}}}{g_{*,0}}
\right)
\left(
\frac{g_{s,0}}{g_{s,{\rm col}}}
\right)^{4/3}
\int d\ln M\,
\frac{d\beta_{\rm wall}}{d\ln M}
\left\langle
\epsilon_{\rm GW}(M,e)\,
{\cal S}\!\left(\frac{f}{f_0(M,e)}\right)
\right\rangle_e .
\end{equation}
The calculation is a reconstruction of the possible stochastic signal from the wall sector. It is not a direct likelihood fit to the PTA data, the NANOGrav band is used as a phenomenological target region.

\begin{figure}[h!]
    \centering
    \includegraphics[width=0.82\textwidth]{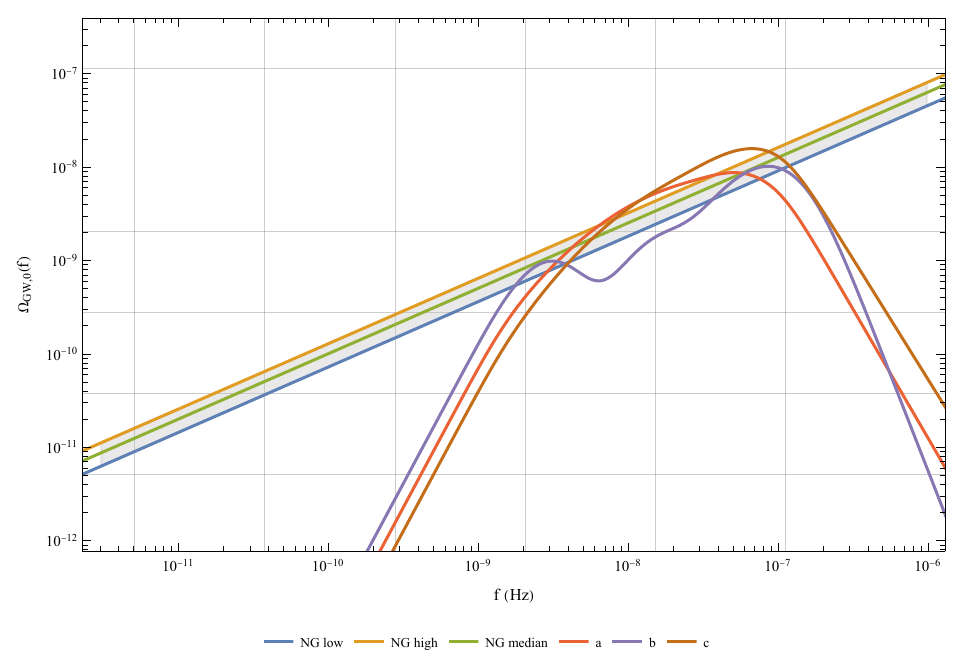}
    \caption{Closed wall gravitational-wave spectra compared with the adopted NANOGrav band. The curves \((a)\), \((b)\), and \((c)\) correspond to the three representative scan points listed in Table~\ref{tab:fixed_lambda_gw_top3}.}
    \label{fig:closed_wall_gw_nanograv}
\end{figure}

We find three representative spectra that enter the adopted NANOGrav band. We label these three solutions as \((a)\), \((b)\), and \((c)\), following the notation used in Fig.~\ref{fig:closed_wall_gw_nanograv}. 
The spectral diagnostics are shown in Table~\ref{tab:fixed_lambda_gw_top3}.
\begin{table}[h!]
\centering
\scriptsize
\begin{tabular}{c|c|c|c|c|c|c}
\hline
Case &
\(f_{\rm peak}\,[{\rm Hz}]\) &
\(\Omega_{{\rm GW},0}^{\rm peak}\) &
\(f_{\rm NG}\,[{\rm Hz}]\) &
\(\Omega_{{\rm GW},0}(f_{\rm NG})\) &
\(N_{\rm band}\) &
\(d_{\rm band}\) \\
\hline
\((a)\) &
\(4.9403\times 10^{-8}\) &
\(8.6701\times 10^{-9}\) &
\(3.9242\times 10^{-9}\) &
\(1.2196\times 10^{-9}\) &
\(25\) &
\(7.9889\times 10^{-3}\) \\
\((b)\) &
\(9.3057\times 10^{-8}\) &
\(1.0050\times 10^{-8}\) &
\(9.3057\times 10^{-9}\) &
\(1.0050\times 10^{-8}\) &
\(25\) &
\(8.7178\times 10^{-3}\) \\
\((c)\) &
\(6.7803\times 10^{-8}\) &
\(1.5672\times 10^{-8}\) &
\(5.3858\times 10^{-9}\) &
\(1.5287\times 10^{-9}\) &
\(24\) &
\(1.5472\times 10^{-3}\) \\
\hline
\end{tabular}
\caption{Three representative gravitational wave spectra. Here \(f_{\rm NG}\) denotes the frequency at which the spectrum is closest to the adopted NANOGrav proxy band, \(N_{\rm band}\) is the number of sampled points lying inside the band, and \(d_{\rm band}\) is the corresponding logarithmic band distance diagnostic.}
\label{tab:fixed_lambda_gw_top3}
\end{table}

The signal reaches peak amplitudes
\begin{equation}
\Omega_{{\rm GW},0}^{\rm peak}
\sim
10^{-8},
\end{equation}
with peak frequencies in the range
\begin{equation}
f_{\rm peak}
\simeq
(4.9-9.3)\times 10^{-8}\ {\rm Hz}.
\end{equation}

This result should be interpreted as a phenomenological existence proof. The present calculation does not claim a unique prediction for the PTA signal, because the abundance of closed walls, the collapse compactness, and the nonsphericity distribution are still treated phenomenologically. A more restrictive test would require a first principles calculation of the closed wall abundance and collapse shape distribution, followed by a direct likelihood comparison with PTA data.

\section{Conclusion}
In this paper we have proposed a novel ALP model which is safe from isocurvature constraint on the CMB scales. The model is based on a complex scalar field with a mexican hat potential coupled to the GB term, which allows to postpone the $U(1)$ symmetry breaking. 

In our analysis, we performed calculations under certain constraints on the parameters of the model, targeting the scenario where the complex Peccei--Quinn-like field does not backreact on inflation. Under our assumptions, the model provides an ultra-light ALP, which can describe fuzzy dark matter. As fuzzy dark matter is now constrained in a wide mass range, it can only constitute a subdominant component of the whole dark matter.

We also discuss the possibility of the ALP serving as dark energy. An interesting feature of this scenario is that dark energy is inhomogeneous on small scales, and can thus help to resolve the Hubble tension. 

The possibility of PBH production is also investigated and was shown to be completely excluded under our assumptions (no backreaction from the symmetry breaking field and effectively single-field inflation), although a more complicated scenarios with multi-field GB-coupled inflation can be explored in future works. Following the constraints on fuzzy dark matter, possible domain walls simply cannot collapse into a black hole. Though, we note that the wall formation itself is highly suppresed due to the low possibility of crossing the axion potential's maximum.

Finally, we estimated the stochastic gravitational wave background produced by the late collapse of closed domain wall configurations. With conservative collapse parameters the signal is too small to be detected. However, after allowing an explicit wall production (with a different parameter choice, unrelated to the fuzzy dark matter scenario) and collapse shape sector, we found an extreme phenomenological branch whose spectrum reaches the NANOGrav band.

We would like to emphasize that the main feature of our model is a scale-dependent axion power spectrum with peak at small scales, which provides a possibility of strong primordial inhomogeneities that could facilitate structure formation in the early Universe. Therefore, we conclude that our main result is the demonstration that a combination of the particle physics beyond the standard model with modified gravity effects could provide non-trivial cosmological scenarios. It is also important that the model is falsifiable, since there are constraints on fuzzy dark matter as well as the observations by DESI, which could potentially provide some constraint on dark energy models.

\section*{Acknowledgements}

We are grateful to V.D. Stasenko and A. Shah for fruitful discussions and their interest in our study. The~research of M.K. was carried out in the Southern Federal University with financial support from the Ministry of Science and Higher Education of the Russian Federation (State contract FENW-2026-0028).

\appendix

\newpage \section*{Appendix A: Suppression of the mixing between radial and phase fluctuations}\label{App_A}

\begin{equation}
S=\int d^4x\sqrt{-g}\left[
\frac{1}{2}R-\frac{1}{8}\xi(\phi)R^2_{\rm GB}
-\frac{1}{2}(\partial\chi)^2-\frac{1}{2}(\partial\phi)^2-\frac{1}{2}\phi^2(\partial\theta)^2
- V_{\rm inf}(\chi)-\frac{\lambda}{4}(f^2-\phi^2)^2
\right],
\label{eq:action}
\end{equation}

During inflation the ALP $\theta$ is effectively shift symmetric and has no potential.

We assume a spatially flat FRW background,
\begin{equation}
ds^2=-dt^2+a(t)^2 d\mathbf{x}^2,
\qquad
\chi=\bar\chi(t),
\qquad
\phi=\bar\phi(t),
\qquad
\theta=\bar\theta(t).
\end{equation}
The homogeneous ALP equation is obtained from $\nabla_\mu(\phi^2\nabla^\mu\theta)=0$ and reads
\begin{equation}
\ddot{\bar\theta}+\left(3H+2\frac{\dot{\bar\phi}}{\bar\phi}\right)\dot{\bar\theta}=0,
\label{eq:theta_bg}
\end{equation}
so that
\begin{equation}
\dot{\bar\theta}(t)\propto a(t)^{-3}\,\bar\phi(t)^{-2}.
\label{eq:thetadot_decay}
\end{equation}
When $\bar\phi$ varies slowly, this implies a rapid decay of the background angular velocity.

In the Gauss--Bonnet sector it is convenient to introduce the parameter
\begin{equation}
\omega \equiv \dot\xi\,H = \xi_{,N}H^2,
\end{equation}
and we will assume the regime in which inflation is effectively single field, meaning that the PQ sector does not backreact on the inflationary background trajectory.
A sufficient condition used in the underlying analysis is
\begin{equation}
|\omega|,\ |\omega_{,N}|\ll \epsilon\ll 1,
\label{eq:omega_small}
\end{equation}
with $\epsilon\equiv -\dot H/H^2$.
This is the same hierarchy that allows one to treat $\chi$ as the inflationary clock while keeping $\phi$ and $\theta$ as spectators for the metric background.

In terms of the ADM variables, the metric is written as
\begin{gather}
g_{00}=-(1+\alpha)^2+a^{-2}\partial_i\beta\,\partial^i\beta,
\qquad
g_{0i}=\partial_i\beta,
\qquad
g_{ij}=a^2\delta_{ij},
\label{eq:metric_alpha_beta}
\end{gather}
and we perturb the PQ fields as
\begin{equation}
\phi(t,\mathbf{x})=\bar\phi(t)+\delta\phi(t,\mathbf{x}),
\qquad
\theta(t,\mathbf{x})=\bar\theta(t)+\delta\theta(t,\mathbf{x}).
\label{eq:field_pert}
\end{equation}
We do not perturb the inflaton, $\delta\chi=0$, consistently with the assumption of effectively single field inflation for the background.

The standard method is to expand the action to second order in perturbations, solve the constraint equations for $\alpha$ and $\beta$, and substitute the solutions back to obtain a reduced quadratic action for the propagating degrees of freedom.
This is the same strategy used in the single field case and its multi field extensions.

For the present purpose we only need the structure of the quadratic action that contains $\alpha$ and $\beta$.
Up to total derivatives, and keeping terms that are quadratic in perturbations and at most linear in $\alpha$ and $\beta$, the PQ sector contributes schematically
\begin{align}
a^{-3}\mathcal{L}^{(2)} \supset\,
&-\Big[3H^2(1-2\omega)-\frac12(\dot{\bar\chi}^2+\dot{\bar\phi}^2+\bar\phi^{\,2}\dot{\bar\theta}^{\,2})\Big]\alpha^2
-(2-3\omega)H\,\alpha\,\frac{\partial^2}{a^2}\beta
\nonumber\\
&+\tfrac{1}{2}\delta\dot\phi^2+\tfrac{1}{2}\delta\phi\frac{\partial_i^2}{a^2}\delta\phi-\tfrac{1}{2}\Big[V_{,\phi\phi}+3\xi_{,\phi\phi}H^4(1-\epsilon)-\dot{\bar\theta}^2\Big]\delta\phi^2-(\dot{\bar\phi}+3\xi_{,\phi}H^3)\alpha\delta\dot\phi
\nonumber\\
&-(V_{,\phi}+3\xi_{,\phi\phi}H^3\dot{\bar\phi}+\bar\phi\dot{\bar\theta}^2)\alpha\delta\phi+(\dot{\bar\phi}+\xi_{,\phi}H^3-\xi_{,\phi\phi}H^2\dot{\bar\phi})\delta\phi\frac{\partial_i^2}{a^2}\beta+\xi_{,\phi}H^2\alpha\frac{\partial_i^2}{a^2}\delta\phi\nonumber\\
&-\xi_{,\phi}H^2\delta\dot\phi\frac{\partial_i^2}{a^2}\beta-\bar\phi^2\dot{\bar\theta}\alpha\delta\dot\theta+\tfrac{1}{2}\bar\phi^2\delta\dot\theta^2+\bar\phi^2\dot{\bar\theta}\delta\theta\frac{\partial_i^2}{a^2}\beta+\tfrac{1}{2}\bar\phi^2\delta\theta\frac{\partial_i^2}{a^2}\delta\theta+2\bar\phi\dot{\bar\theta}\delta\phi\delta\dot\theta~.
\label{eq:L2_structure}
\end{align}

The Gauss--Bonnet modification now enters the constraint sector not only through the coefficients of $\alpha^2$ and $\alpha\,\partial^2\beta$, but also through the explicit $\xi_{,\phi}$ and $\xi_{,\phi\phi}$ terms that mix $\delta\phi$, $\delta\dot\phi$, $\alpha$, and $\beta$. For convenience, let us define
\begin{equation}
\mathcal{A}_0\equiv 3H^2(1-2\omega)-\tfrac{1}{2}\big(\dot{\bar\chi}^2+\dot{\bar\phi}^2\big),
\qquad
D\equiv (2-3\omega)H,
\label{eq:A0_D_def}
\end{equation}
as well as
\begin{equation}
R\equiv \dot{\bar\phi}+\xi_{,\phi}H^3-\xi_{,\phi\phi}H^2\dot{\bar\phi},
\qquad
S\equiv \xi_{,\phi}H^2,
\qquad
P\equiv \dot{\bar\phi}+3\xi_{,\phi}H^3,
\qquad
Q_0\equiv V_{,\phi}+3\xi_{,\phi\phi}H^3\dot{\bar\phi}.
\label{eq:RSQ_def}
\end{equation}
Here $\mathcal{A}_0$ is the part of the coefficient of $\alpha^2$ that is independent of $\dot{\bar\theta}^2$. This is sufficient because any insertion of the $\bar\phi^2\dot{\bar\theta}^2$ piece of $\mathcal{A}$ into the $\delta\phi_k$ source would generate terms of order $\dot{\bar\theta}^3\delta\phi_k$, which we discard.

Varying \eqref{eq:L2_structure} with respect to $\beta$ gives the momentum constraint. Since $\beta$ enters only through $\partial^2\beta/a^2$, one finds
\begin{equation}
-(2-3\omega)H\,\alpha+R\,\delta\phi-S\,\delta\dot\phi+\bar\phi^{\,2}\dot{\bar\theta}\,\delta\theta=0.
\end{equation}
Therefore the lapse perturbation is
\begin{equation}
\alpha
=\frac{R\,\delta\phi-S\,\delta\dot\phi+\bar\phi^{\,2}\dot{\bar\theta}\,\delta\theta}{D}.
\label{eq:alpha_solution_general}
\end{equation}

Varying \eqref{eq:L2_structure} with respect to $\alpha$ gives the Hamiltonian constraint,
\begin{align}
-2\mathcal{A}\,\alpha&-(2-3\omega)H\,\frac{\partial^2}{a^2}\beta
-(\dot{\bar\phi}+3\xi_{,\phi}H^3)\delta\dot\phi
-\big(V_{,\phi}+3\xi_{,\phi\phi}H^3\dot{\bar\phi}+\bar\phi\dot{\bar\theta}^2\big)\delta\phi \nonumber \\ 
&+\xi_{,\phi}H^2\frac{\partial^2}{a^2}\delta\phi
-\bar\phi^{\,2}\dot{\bar\theta}\,\delta\dot\theta=0.
\end{align}
Keeping only terms that can contribute up to order $\dot{\bar\theta}\,\delta\phi_k$ in the ALP source, this becomes
\begin{equation}
-\frac{\partial^2}{a^2}\beta=
\frac{
2\mathcal{A}_0\,\alpha
+P\,\delta\dot\phi
+Q_0\,\delta\phi
-\xi_{,\phi}H^2\frac{\partial^2}{a^2}\delta\phi
+\bar\phi^{\,2}\dot{\bar\theta}\,\delta\dot\theta
}{D}.
\label{eq:beta_solution_general}
\end{equation}
Again, the omitted pieces would only generate terms of order $\dot{\bar\theta}^3\delta\phi_k$ or higher in the final ALP equation.

Substituting \eqref{eq:alpha_solution_general} and \eqref{eq:beta_solution_general} into the ALP equation and keeping all terms up to linear order in $\dot{\bar\theta}$ on the right-hand side and up to order $\dot{\bar\theta}^2$ in the ALP self-sector, one finds for a Fourier mode $\delta\theta_k$
\begin{align}
\delta\ddot\theta_k
&+\left(3H+2\frac{\dot{\bar\phi}}{\bar\phi}\right)\delta\dot\theta_k
+\left(\frac{k^2}{a^2}+m^2_{\theta,{\rm eff}}\right)\delta\theta_k \nonumber \\
&=
-\frac{\xi_{,\phi}H\,\dot{\bar\theta}}{2-3\omega}\,\delta\ddot\phi_k
+\dot{\bar\theta}\,\mathcal{C}_{\dot\phi}\,\delta\dot\phi_k
+\dot{\bar\theta}\,\mathcal{C}_{\phi}\,\delta\phi_k
+\dot{\bar\theta}\,\mathcal{C}_{k}\,\frac{k^2}{a^2}\delta\phi_k
+\mathcal{O}\!\left(\dot{\bar\theta}^{\,3}\delta\phi_k\right),
\label{eq:delta_theta_final}
\end{align}
where
\begin{align}
\mathcal{C}_{\dot\phi}
&=
-\frac{2}{\bar\phi}
+\frac{R-\dot S}{D}
+\frac{S(\dot D+2\mathcal{A}_0)}{D^2}
-\frac{P}{D},
\label{eq:Cdotphi}
\\[3pt]
\mathcal{C}_{\phi}
&=
\frac{2\dot{\bar\phi}}{\bar\phi^2}
+\frac{\dot R}{D}
-\frac{R(\dot D+2\mathcal{A}_0)}{D^2}
-\frac{Q_0}{D},
\label{eq:Cphi}
\\[3pt]
\mathcal{C}_{k}
&=
-\frac{\xi_{,\phi}H^2}{D}
=
-\frac{\xi_{,\phi}H}{2-3\omega}.
\label{eq:Ck}
\end{align}
The effective ALP mass term is
\begin{equation}
m^2_{\theta,{\rm eff}}
=
\bar\phi^{\,2}\dot{\bar\theta}^{\,2}
\left[
\frac{3H}{D}
+\frac{\dot D+2\mathcal{A}_0}{D^2}
\right]
+\mathcal{O}\!\left(\dot{\bar\theta}^{\,4}\right).
\label{eq:mthetaeff_exact}
\end{equation}

It is worth stressing the structure of \eqref{eq:delta_theta_final}. After the consistent truncation described above, every source term involving $\delta\phi_k$ is proportional to a single power of $\dot{\bar\theta}$. There are no surviving terms of order $\dot{\bar\theta}^2\delta\phi_k$ on the right-hand side. Such contributions first appear only at order $\dot{\bar\theta}^3\delta\phi_k$, which has been discarded.

Thus the corrected calculation still leads to the same physical conclusion. If the background angular velocity is negligible, then the right-hand side of \eqref{eq:delta_theta_final} vanishes at linear order, and $\delta\theta_k$ evolves independently of $\delta\phi_k$. In particular, using the background relation
\begin{equation}
\dot{\bar\theta}\propto a^{-3}\bar\phi^{-2},
\label{eq:thetadot_decay_again}
\end{equation}
the mixing terms are rapidly redshifted away provided that $\bar\phi$ does not grow exponentially.

A simple measure of the mixing is again
\begin{equation}
\epsilon_{\rm mix}(t)\equiv \left|\frac{\dot{\bar\theta}}{H}\right|.
\end{equation}
The suppression argument requires that $\bar\phi$ vary slowly on Hubble time scales, namely
\begin{equation}
\left|\frac{\dot{\bar\phi}}{H\bar\phi}\right|\ll 1,
\qquad
\left|\frac{\ddot{\bar\phi}}{H^2\bar\phi}\right|\ll 1.
\label{eq:SR_phi_cond}
\end{equation}
Under \eqref{eq:SR_phi_cond}, Eq.~\eqref{eq:thetadot_decay_again} implies that the source terms on the right-hand side of \eqref{eq:delta_theta_final} are transient and quickly become negligible.

In the slow-roll regime, after using the background ALP equation, the slow variation condition, the Gauss--Bonnet slow-roll condition, and Eq.~\eqref{eq:SR_phi_cond}, the effective mass term \eqref{eq:mthetaeff_exact} reduces to the same leading contribution as in the ALP-only treatment,
\begin{equation}
m^2_{\theta,{\rm eff}}\simeq 3\bar\phi^{\,2}\dot{\bar\theta}^{\,2}.
\end{equation}
Then Eq.~\eqref{eq:delta_theta_final} becomes
\begin{equation}
\delta\ddot\theta_k
+3H\delta\dot\theta_k
+\left(\frac{k^2}{a^2}+3\bar\phi^{\,2}\dot{\bar\theta}^{\,2}\right)\delta\theta_k
\simeq 0,
\end{equation}
once the transient $\delta\phi_k$ source terms have redshifted away. Equivalently,
\begin{equation}
u_{k,NN}+u_{k,N}+\Big[\frac{k^2}{a^2H^2}-2+3\bar\phi^{\,2}\bar\theta_{,N}^2(N_{\rm in})e^{6(N_{\rm in}-N)}\Big]u_k\simeq 0,
\qquad
u_k\equiv a\bar\phi\,\delta\theta_k.
\end{equation}
Provided
\begin{equation}
\bar\phi^{\,2}(N_{\rm in})\bar\theta_{,N}^2(N_{\rm in})\ll 1,
\end{equation}
the last term is quickly suppressed so that the standard quasi-de Sitter solution for $\delta\theta_k$ can still be used.

Therefore, after correcting the quadratic action and consistently substituting the new $\alpha$ and $\beta$ into the ALP equation, the conclusion remains unchanged. Radial fluctuations can source the ALP only through terms proportional to the background angular velocity $\dot{\bar\theta}$. Since $\dot{\bar\theta}$ decays rapidly during inflation when $\bar\phi$ varies slowly, radial fluctuations do not affect the ALP perturbation at linear order after a short transient. This supports the use of the decoupled $\delta\theta_k$ mode function in the stochastic ALP construction.

\clearpage

\bibliography{Bibliography.bib}{}
\bibliographystyle{utphys.bst}

\end{document}